\documentclass[twocolumn,showpacs,preprintnumbers,amsmath,amssymb]{revtex4}

\usepackage{subfigure}
\usepackage[dvips]{graphicx}
\usepackage{dcolumn}
\usepackage{bm}
\newcommand{\eq}{\begin{equation}}
\newcommand{\eqq}{\end{equation}}
\newcommand{\dd}{\text{d}}

\begin{document}

\title{Polymer-induced tubulation in lipid vesicles}

\author{F. \surname{Campelo}}
\email{campelo@ecm.ub.es}
\author{A. \surname{Hern\'{a}ndez--Machado}}%
\affiliation{Departament d'Estructura i Constituents de la Mat\`{e}ria,\\
Facultat de F\'{\i}sica, Universitat de Barcelona\\
  Diagonal 647, E-08028 Barcelona, Spain}

\date{\today}

\begin{abstract}

A mechanism of extraction of tubular membranes from a lipid vesicle is presented. A concentration gradient of anchoring amphiphilic polymers generates tubes from bud-like vesicle protrusions. We explain this mechanism in the framework of the Canham-Helfrich model. The energy profile is analytically calculated and a tube with a fixed length, corresponding to an energy minimum, is obtained in a certain regime of parameters. Further, using a phase-field model, we corroborate these results numerically. We obtain the growth of tubes when a polymer source is added, and the bud-like shape after removal of the polymer source, in accordance with recent experimental results.

\end{abstract}

\maketitle

\textit{Introduction.--} Vesicle shape transformations, like tubulation, play an essential role in cellular transport.
Both energy and matter are transferred continuously throughout the membrane, a lipid bilayer with proteins and other macromolecules anchored on it, which surrounds the cell and most of its internal organelles \cite{alberts}.
As part of cellular dynamic processes, these membranes adopt different shapes in order to exchange matter with their surroundings. Many possibilities appear here, from budding and eventual fission of small transport vesicles \cite{farsad03} to formation of large tethers connecting distant organelles, as in the Golgi apparatus and the endoplasmic reticulum \cite{rapoport06}, or even between different cells~\cite{rustom04}.

\begin{figure}[ht!]
   \centering
   \includegraphics[width=8cm]{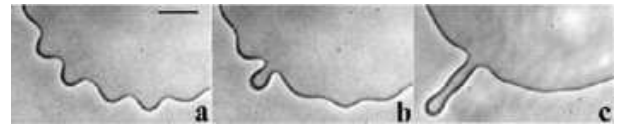}
   \caption{Tube formation in a highly oblate vesicle after local addition of multianchor polymer. Experimental images from Ref.~\cite{tsafrir03}.}
   \label{fig:exp.tubes}
\end{figure}

The formation of these tethers can be driven by applying a point-like force to the membrane \cite{powers02,derenyi02}. Understanding the nature of this force is of major importance. There are different mechanisms leading to such a tubulation phenomenon, as for instance the growing of microtubules pushing the membrane from inside \cite{libchaber97}, and the extrusion due to a hydrodynamic flow \cite{nassoy06}. Other works have experimentally studied the force generated by molecular motors pulling membrane tubes in vitro \cite{roux02} and by optical tweezers \cite{koster05}. Tsafrir \emph{et al.} studied the tubulation induced in highly oblate vesicles by the anchoring of amphiphilic polymers \cite{tsafrir03} without any directed force. In those experiments, macromolecules containing hydrophobic groups were administered in the surroundings of a giant oblate vesicle. Those molecules diffused in the bulk and eventually anchored the membrane inducing a local spontaneous curvature by the mechanism of hydrophobic insertion \cite{kozlov06}, leading to the formation of one or several buds. Even, under certain circumstances, those buds can grow into long tubular structures (see Fig.~\ref{fig:exp.tubes}).

Here, motivated by those experimental results \cite{tsafrir03}, we present a theoretical treatment of a novel mechanism of tube extraction, due to the generation of spontaneous curvature by anchoring macromolecules which are distributed along a gradient of concentration, maintained by a source.
We show that the elongation of a bud into a tube may be, under certain circumstances, energetically favorable when a polymer concentration gradient is present. We obtain that these tubes do not grow indefinitely, but stop at a certain length.
In the first part of this Letter, we analyze the problem in the framework of the Canham-Helfrich model \cite{canham70,helfrich73} with a simple geometry and a stationary linear concentration profile, in order to solve analytically and to understand qualitatively the tube formation. In the second part, we use a phase-field model for the bending energy \cite{campelo_prl_07}, coupled with a stationary polymer concentration profile, to study the problem numerically.

\begin{figure}[ht!]
	\centering
	\includegraphics[width=7.5cm,bb=0 0 459 224]{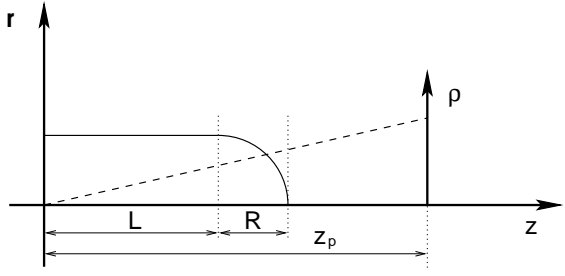}
	\caption{Geometrical sketch of the system: a cylindrical tube of radius $R$ and length L, with a hemispherical cap of radius $R$. A linear polymer concentration gradient is $\rho(z)$ is also outlined in the figure (dashed line).}
	\label{fig:tubepolygeom}
\end{figure}

\textit{Theoretical treatment.--} Let us assume that a hemispherical bud of radius $R$ is already formed out from a giant vesicle. This bud has not to be necessarily stable, appearing for instance just as a vesicle rim fluctuation \cite{tsafrir03}. The mother vesicle is large enough to be considered as a lipid reservoir during the whole extraction process, so we can study the isolated tube on its own. In addition, there are amphiphilic macromolecules (e.g. polymers as in \cite{tsafrir03}) in the volume outside the vesicle. We assume here, in order to find analytical estimations for the extraction of membrane tubes due to an inhomogeneous polymer concentration in the bulk, that these macromolecules are applied in a line located at a distance $z_p$ from the mother vesicle (see Fig.~\ref{fig:tubepolygeom}) and follow a linear stationary profile for this concentration, $\rho(\mathbf{r},t)=\rho_0\, \left(1-|z/z_p-1|\right)$, where $\mathbf{r_{\text{appl}}}=(0,0,z_p)$.

The bending energy of a membrane treated as an elastic sheet is given by the Canham-Helfrich Hamiltonian \cite{canham70,helfrich73}
\eq\label{bendingenergyH}
E=\int_{S}{\left(\frac{\kappa}{2}\left[c_1(\mathbf{r})+c_2(\mathbf{r})-c_0(\mathbf{r})\right]^2+\sigma\right)\dd^2 A},
\eqq
where $c_1$ and $c_2$ are the principal curvatures of the surface, $c_0$ the spontaneous curvature and $\kappa$ the bending modulus. Since the tube is connected to a lipid reservoir, we have also included a surface tension term, where $\sigma$ is the surface tension of the membrane. A pressure-like term could also be introduced, but its effects are subleading \cite{powers02}. Using a cylindrical tube with a hemispherical cap, as sketched in Fig.~\ref{fig:tubepolygeom}, Eq.~\ref{bendingenergyH} can be written as
\begin{eqnarray}
\label{bendingenergy}
E&=&\int_{S_L}{\left(\frac{\kappa}{2}\left[\frac{1}{R}-c_0(\mathbf{r})\right]^2+\sigma\right)\dd^2  A}\nonumber\\
&+&\int_{S_R}{\left(\frac{\kappa}{2}\left[\frac{2}{R}-c_0(\mathbf{r})\right]^2+\sigma\right)\dd^2 A},
\end{eqnarray}
where $S_L$ and $S_R$ denote the cylindrical and the hemispherical parts of the tube (Fig.~\ref{fig:tubepolygeom}) with areas $A_L$ and $A_R$, respectively. The spontaneous curvature of the membrane is coupled with the local concentration since the hydrophobic anchor groups of the polymer tend to insert themselves into the bilayer, acting thus as a wedge \cite{leibler86}. A linear coupling, as stated previously in a similar system \cite{tsafrir01,campelo_prl_07}, has shown to be justified, the spontaneous curvature field outside the vesicle being $c_0(z)=\bar{c}_0\, \rho(z)$.

Therefore, in the case where the polymer gradient in space is set to be linear, we get for the tube energy Eq.~(\ref{bendingenergy})
\begin{eqnarray}
\label{energy.linear}
\frac{E}{\kappa \pi}&=&
\frac{1}{3}\xi ^2 L^3
+\xi\left(\xi-1\right) L^2
+\left(\frac{\pi  \xi ^2}{2}-4 \xi +2 \bar{\sigma} +1\right) L
\nonumber\\
&&
+\left(\frac{2 \xi ^2}{3}-\pi  \xi +2 \bar{\sigma} +4\right)
\end{eqnarray}
where $R=1$ sets the length scale, $\bar{\sigma}=\sigma/\kappa$, and we defined $\xi=\bar{c}_0\,\rho_0/z_p$, as the slope of the linear spontaneous curvature profile.

\begin{figure}[ht!]
	\centering
	\includegraphics[width=7.5cm]{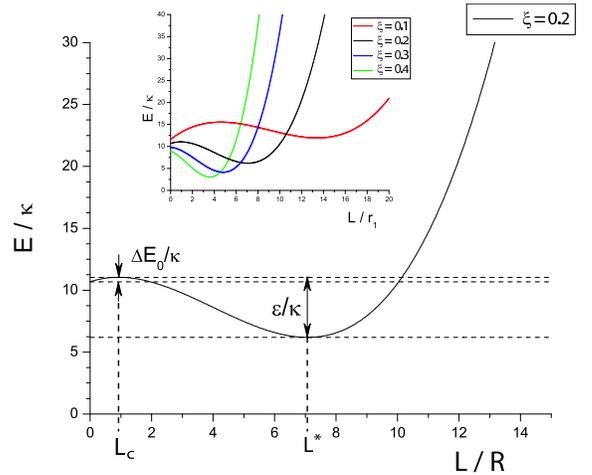}
	\caption{Energy vs. tube length for different values of the rate of polymer molecules added to the system, i.e. the slope of the linear polymer profile $\xi$ (inset). Initial buds need to overcome an energy barrier $\Delta E_0$ to elongate into tubes for a certain range of values of $\xi$. Then, the elongated tube needs to overcome another barrier $\epsilon$ in order to be reabsorved by the mother vesicle. The surface tension here is negligible.}
	\label{fig:enlinall}
\end{figure}

\begin{figure}[ht!]
	\centering
	\includegraphics[width=7cm,bb=0 0 263 204]{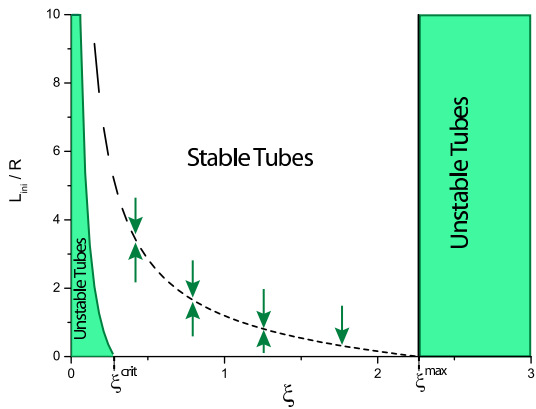}
	\caption{Stability diagram of tube formation as a function of the initial tube of length $L_{\text{ini}}$ ($L_{\text{ini}}=0$ corresponds to a bud), and the dimensionless slope of the spontaneous curvature gradient $\xi$. The dashed line correspond to the stable tube length, $L^*$, for a given concentration gradient.}
	\label{fig:StabilityDiagram}
\end{figure}

In Fig.~\ref{fig:enlinall}~(inset) we show how the tube energy Eq.~(\ref{energy.linear}) looks like as a function of the length $L$ of the tube for different slopes of the spontaneous curvature profile, $\xi$, when the tension $\sigma$ is negligible. Note that the energy Eq.~(\ref{energy.linear}) is cubic in the tube length. The energy extremes correspond to two equilibrium lengths: one stable length corresponding to an energy minimum, $L^*=1/\xi-1+\sqrt{2/\xi-\pi/2+1}$; and another smaller length being unstable, $L_c=1/\xi-1-\sqrt{2/\xi-\pi/2+1}$. The larger the slope of the linear concentration profile is, the smaller these lengths are. For $\xi>\xi^{\text{crit}}=4/ \pi (1-\sqrt{1-\pi/8})\simeq 0.28$ (see Fig.~\ref{fig:StabilityDiagram}), theres is only a local minimum and no local maximum of the energy for positive lengths (Fig.~\ref{fig:enlinall}). This means that the initial bud becomes unstable against the formation of a tube. For smoother concentration gradients ($\xi<\xi^{\text{crit}}$), the initial bud, in order to grow up to its preferred length $L^*$, needs to cross an energy barrier $\Delta E_0$, or to start with a certain initial length larger than $L_c$ (see Fig.~\ref{fig:enlinall}). For very steep slopes of the concentration gradient $\xi>\xi^{\text{max}}=4/ \pi (1+\sqrt{1-\pi/8})\simeq 2.27$, the local energy minimum at $L^*$ disappear, and no stable tubes can be found (see Fig.~\ref{fig:StabilityDiagram}).

For an initially formed bud-like fluctuation ($L_\text{ini}=0$ in Fig.~\ref{fig:StabilityDiagram}), when we increase the slope of the spontaneous curvature, $\xi$, the bud cannot grow unless the critical slope, $\xi^{\text{crit}}$, is reached, and then a finite-length tube is formed out (dashed line in Fig.~\ref{fig:StabilityDiagram}). This transition is discontinuous in the value of the stable length of the tube. In addition, we see that the tube length is maximum for $\xi^{\text{crit}}$, decreasing for steeper profiles, $\xi>\xi^{\text{crit}}$. Eventually, at $\xi^{\text{max}}$ the equilibrium tube length vanishes, and buds become again unstable. This transition is therefore continuous.

For non-vanishing but relatively small values of the surface tension, $\sigma$, the stable tube length decreases as $L^*(\bar{\sigma})=L^*(0)-\bar{\sigma}/(\xi^2\, \sqrt{1-\pi/2+2/\xi)})$, where $L^*(0)$ is the tube length for zero tension. For tensions larger than a critical tension $\bar{\sigma}_c=4-\pi/2$, bud to tube transition dissapear, altough stable tubes may be formed out from a tube with a finite given length.

\textit{Numerical model.--} In the experiments by Tsafrir \emph{et al.} \cite{tsafrir03}, polymer molecules diffuse from a source and eventually anchor the membrane, inducing a local spontaneous curvature which modifies the vesicle shape. In order to study dynamically this process, we present here a numerical model dealing with a stationary polymer profile and the Canham-Helfrich energy without any assumption on the tube geometry. A systematic study of other stationary and non-stationary concentration profiles, and the diffusion of the polymer molecules and their anchorage on the membrane, is beyond the scope of this Letter, and will be presented elsewhere \cite{campelo_phd}.

We use a phase-field method \cite{campelo06,campelo_epjst_07,campelo_prl_07,misbah03,duliuwang04,voigt07} to numerically deal with the bending energy of the membrane. Within this approach, the Canham-Helfrich Hamiltonian Eq.~(\ref{bendingenergyH}) is written as a dynamic function of a field, $\phi$, whose level-set $\{\mathbf{x}:\phi(\mathbf{x})=0\}$ locates the membrane position at each time. The dynamic equation for the phase-field, and hence for the membrane shape, is \cite{campelo_prl_07}
\begin{eqnarray}\label{dyn.eqn}
\frac{\partial \phi}{\partial t}&=&2\bm{\nabla}^2\Big\{\left(3\phi^2-1+2\epsilon C_0(\rho(\bm{x}))\, \phi\right)\Phi[\phi,\rho]\nonumber\\
&-&\epsilon^2\bm{\nabla}^2\Phi[\phi,\rho]+\epsilon^2\tilde{\sigma}\bm{\nabla}^2 \phi
\Big\},
\end{eqnarray}
where $\Phi[\phi(\bm{x}),\rho(\bm{x})]=(\phi^2-1)\left(\phi-\epsilon\,C_0(\rho(\bm{x}))\right)-\epsilon^2\bm{\nabla}^2\phi(\bm{x})$, $C_0(\rho(\bm{x}))=\bar{c}_0\,\rho(\mathbf{x})$ is the spontaneous curvature induced by the local concentration $\rho(\mathbf{x})$ of polymer molecules anchored on the membrane, and $\epsilon$ is a small parameter related to the width of the interface. This dynamic equation includes a tube surface tension $\tilde{\sigma}$, and conserves locally the inner volume, due to the use of a relaxational model-B-like conserved dynamics.

The assumption we did before of a linear stationary polymer profile is relaxed at this point. Since our aim in this Letter is to show how an inhomogeneous stationary polymer concentration profile is a possible mechanism of tube formation, we are going to use in our numerical treatment a Gaussian stationary profile such as $\rho(\bm{x})=\rho_0/\Sigma\sqrt{2\pi} \exp{\left(-\left|\bm{x}-\bm{x}_{\text{appl}}\right|^2/2 \Sigma^2\right)}$, where $\Sigma$ is the standard deviation, related with the width of the distribution, and $\rho_0$ is the total number of polymer molecules introduced in the system. Our problem is then reduced to numerically solve Eq.~(\ref{dyn.eqn}) using this stationary polymer concentration for different sets of initial conditions and parameters.  We used a standard finite-difference scheme for the spatial discretization and an Euler method for the time-derivatives \cite{campelo06}. In all the results shown in this Letter we used the value of the small parameter, $\epsilon$, equal to the mesh size of the lattice. The results are robust under variations of this parameter. Both the time and space discretizations are chosen in order to satisfy the Courant-Friedrichs-Lewy stability criterion \cite{bertsekas}.

The parameters which are relevant in these simulations in order to study the growth and the shrinkage of tubes are: the length of the initial tube, and the characteristics of the polymer gradient profile.
We fixed the position of the polymer source to be $z_p=7.5\, R$, where $R$ is the radius of the initial tube formed by a cylinder and a hemispherical cap. Due to the axisymmetry of the problem, the integration is performed in a two-dimensional lattice, whose size is $80\times 20$ throughout this Letter.

\begin{figure}[ht]
   \centering
   \includegraphics[width=8.5cm]{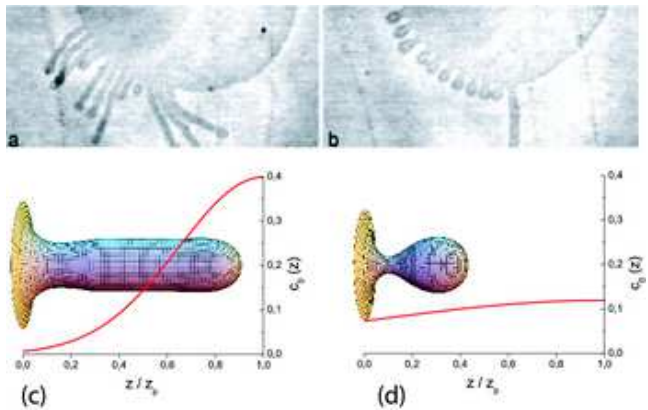}
   \caption{Tubes extruded from a vesicle by a non-homogeneous polymer profile. Comparison between the experimental results from Ref.~\cite{tsafrir03} (a,b) and the phase-field integrations (c,d). For short times (small standard deviation $\Sigma$) long tubes are obtained (a,c), and for long time (wider distributions) buds appear (b,d). The resulting profile for the spontaneous curvature is shown for the phase-field integrations. We choose $\xi=0.3$, $\Sigma=3.5 R$ (c), and $\xi=0.3$, $\Sigma=10R$ (d).}
   \label{fig:sim.tubes}
\end{figure}

\textit{Numerical results and discussion.--} In Fig.~\ref{fig:exp.tubes} we show three snapshots of the experimental results from Ref.~\cite{tsafrir03}, where a fluctuating oblate vesicle undergoes a shape instability which forms an initial bud and eventually grows into a tube, due to the anchorage of amphiphilic polymer molecules. The presence of the polymer molecules enhance fluctuation of the vesicle rim (Fig.~\ref{fig:exp.tubes}(a)). One can divide the dynamics of this process in four different regimes. The first one is the formation of the initial bud and suppression of rim fluctuations (Fig.~\ref{fig:exp.tubes}(b)). The second regime is when the tube starts to grow up to a certain length (Fig.~\ref{fig:exp.tubes}(c)). Then, in the third regime the polymer source is shut down (Fig.~\ref{fig:sim.tubes}(a)) and tubes shrink reaching a metastable bud-like shape (Fig.~\ref{fig:sim.tubes}(b)). In the last regime, they eventually disappear reabsorbed by the mother vesicle. Within our model, we find, as shown in Fig.~\ref{fig:sim.tubes}(c), the shape of a large tube, which is qualitatively in good agreement with the experimental results. For a large range of values of the standard deviation, $\Sigma$, the tube length we get is essentially the same. In other words, when the source of polymer molecules is shut down, the concentration profile gets stretched as time goes by, but the length of the stable tube continues being the same.

Then, after some time, the polymer molecules are more homogeneously distributed, and a new configuration of short length is found (Fig.~\ref{fig:sim.tubes}(d)). These buds are also in agreement with those found by Tsfarir \emph{et al.} \cite{tsafrir03}, and we can quantitatively compare them by measuring the ratio between their width and length to be around $0.3$.

In other recent experiments, Roux \emph{et al.} \cite{roux06} generated dynamin-coated membrane tubes. They observed the growth of these tubes by the addition of this protein. Further addition of GTP molecules to these tubes induces fission by a conformational change of dynamins. Altough our model is capable of explaining the initial tubulation by the generation of membrane curvature by dynamin, it does not consider topological changes such as fission. An extension of our model including the Gaussian curvature term in the Canham-Helfrich Hamiltonian eq.~(\ref{bendingenergyH}) will be published elsewhere \cite{campelotravasso}.

\textit{Conclusions.--} 
In this Letter we showed how a nonhomogeneous concentration of amphiphilic molecules leads to tube formation from vesicles. The mechanism is explained in the framework of the Canham-Helfrich model for the membrane energy, where the macromolecules anchored on the membrane induce a local spontaneous curvature. First, we analyzed the energetics of such a situation with a simplified geometrical scheme and a linear profile for the concentration, and showed analytically under which conditions a tube may be formed from a vesicle. Afterwards, we used a phase-field model to numerically calculate the shape of tubes under no geometric assumptions with a Gaussian profile for the polymer concentration. These numerical treatment led us to find the tube shape, in agreement with those found experimentally by Tsafrir \emph{et al.} \cite{tsafrir03}.
Two qualitatively different stable shapes are found depending on the characteristics of the polymer distribution: long tubes for initially narrow polymer distributions, and short buds for wider ones.


We are indebted to Joel Stavans for critical reading of the manuscript and
for kindly providing us with his experimental results. We acknowledge financial support of the Direcci\'{o}n General de Investigaci\'{o}n under project No. FIS2006-12253-C06-05. F.C. also thanks Ministerio de Educaci\'{o}n y Ciencia (Spain) for financial support.

\bibliography{bib}
\end{document}